# A new equation of state for dark energy

Dragan Slavkov Hajdukovic[1]
PH Division CERN
CH-1211 Geneva 23;
dragan.hajdukovic@cern.ch
[1]On leave from Cetinje; Montenegro

**Abstract**
In the contemporary Cosmology, dark energy is modelled by a perfect fluid having a very simple equation of state; pressure is proportional to dark energy density. As an alternative, I propose a more complex equation of state, with pressure being function of three variables: dark energy density, matter density and the size of the Universe. One consequence of the new equation is that, in the late-time Universe, cosmological scale factor is linear function of time; while the standard cosmology predicts an exponential function. The new equation of state allows attributing a temperature to the physical vacuum; a temperature proportional to the acceleration of the expansion of the Universe. The vacuum temperature decreases with the expansion of the Universe, approaching (but never reaching) the absolute zero.

## 1. Introduction

As well known, the cosmological principle (i.e. the statement that at any particular time the Universe is isotropic about every point) determines the Friedman-Robertson-Walker metric

$$ds^2 = c^2 dt^2 - R^2(t)\left[\frac{dr^2}{1-kr^2} + r^2(d\theta^2 + \sin^2\theta\, d\phi^2)\right]; \quad k = +1, -1, 0 \quad (1)$$

where $k = +1$; $k = -1$; $k = 0$ correspond respectively to closed, open and flat Universe.

The dynamics of the above space-time geometry is entirely characterised by the scale factor $R(t)$. In order to determine the function $R(t)$, the Einstein equation $G_{\mu\nu} = -(8\pi G/c^4)T_{\mu\nu}$ must be solved. While the Einstein tensor $G_{\mu\nu}$ is determined by metric (1); we need a model for the energy-momentum tensor ($T_{\mu\nu}$) of the content of the Universe. In view of homogeneity and isotropy of the Universe, a reasonable approximation is to assume the energy-momentum tensor of a perfect fluid; characterised at each point by its proper density $\rho$ and the pressure $p$ in the instantaneous rest frame. Assuming that the cosmological fluid in fact consists of several distinct components (for example, matter, radiation and the vacuum) the final results are cosmological field equations, which may be written in the form:

$$\ddot{R} = -\frac{4\pi G}{3}R\sum_n\left(\rho_n + \frac{3p_n}{c^2}\right) \quad (2)$$

$$\dot{R}^2 = \frac{8\pi G}{3}R^2\sum_n\rho_n - kc^2 \quad (3)$$

However, it is still not enough. In order to solve cosmological field Equations (2) and (3), we need equation of state for every component of the cosmological fluid. The most used equation of state, relating pressure and energy density is:

$$p_n = w_n\rho_n c^2 \quad (4)$$

where the equation-of state-parameter $w_n$ is a constant.

Density $\rho_n$ of a fluid, satisfying the above equation of state, transforms according to the power-law

$$\rho_n = \rho_{n0}\left(\frac{R_0}{R}\right)^n \quad (5)$$

where, as usually, index $0$ denotes the present day value.

Radiation, matter and dark energy (when identified with cosmological constant) are modelled respectively with $w_r = 1/3$ (*i.e. n = 4*); $w_m = 0$ (*i.e. n = 3*) and $w_\Lambda = -1$ (*i.e. n = 0*). In fact, matter is attributed a non-zero density $\rho_m$ and zero pressure $p_m = 0$; that's why sometimes we call it presureless matter. Dark energy is characterized with $w_\Lambda = -1$, i.e., a constant energy density $\rho_\Lambda c^2$ and a constant negative pressure $p_\Lambda = -\rho_\Lambda c^2$. Without dark energy, a purely matter Universe would collapse one day. With dark energy, the actual Universe is in a state of the accelerated expansion. It is amusing that such a simple equation of state as Eq.(2), correctly describes the main observed features of our Universe.

More general, we may drop assumption that the vacuum energy density is constant and denote the corresponding vacuum energy density and pressure by $\rho_v c^2$ and $p_v$ (instead of $\rho_\Lambda c^2$ and $p_\Lambda$). I will use this notation in the rest of the paper. By the way let me note, that in spite of the cosmological constant problem, the words "vacuum energy" and "dark energy" are used as synonyms in the present paper.

## 2. A conjectured equation of state for dark energy

It was recently noticed [1], that dark energy density ($\rho_{de}$) of the Universe, at least when the observed numerical value is in question, satisfies the following, strikingly simple and elegant equation:

$$\rho_{de} = \frac{A}{2\pi} \frac{\hbar}{c} \frac{\ddot{R}}{\lambda_\pi^3} \qquad (6)$$

where $A \approx 2$ is a dimensionless constant; $\lambda_\pi = h/m_\pi c$ is Compton wavelength of a pion and $\ddot{R}$ the acceleration of the expansion of the Universe determined by the cosmological equation (2). The equation (6) may be just a numerical coincidence; but in the present paper it is considered as a fundamental relation.

For simplicity let's consider the cosmological field equations (2) and (3) with only two non-zero densities: matter density $\rho_m$ and density $\rho_v \equiv \rho_{de}/c^2$ corresponding to dark energy density (6). It is easy to check that simultaneous validity of equations (2) and (6) is possible only if dark energy has the following equation of state:

$$p_v = -\frac{1}{3}\left\{\frac{\rho_m}{\rho_v} + 1 + \frac{3}{4}\frac{\lambda_\pi^3}{l_{Pl}^2}\frac{1}{R}\right\}\rho_v c^2 \equiv w_{eff}\rho_v c^2 \qquad (7)$$

where $l_{Pl}$ is Planck length. Hence, my conjecture is that the old equation of state (4) must be replaced by the new one (7); with inevitable major consequences in Cosmology, The equation (7) deserves the following comment: while the conjecture (7) and Eq.(4) are very different, they may be considered as identical in the present day Universe. In fact, using the best known numerical values [2] for $\rho_{m0}$, $\rho_{v0}$ and $R_0$, Eq.(7) leads to $p_{v0} \approx -0.996\rho_{v0}c^2$ (i.e. $w_{eff} \approx -0.996$) what is extremely close to the choice $w_\Lambda = -1$ in Eq.(4). However for large $R$ i.e. $R >> R_0$, $p_v$ approaches the maximum $-\rho_v c^2/3$ (i.e. $w_{eff} = -1/3$); hence the future of the Universe is quite different.

## 3. Evolution of dark energy density

In order to find how vacuum energy density evolves with the cosmological scale factor $R(t)$ it is appropriate to start with the second cosmological equation (3). The first step is to differentiate Eq.(3) with respect to time and after differentiation to introduce $\ddot{R}$ (determined by Eq.(6)) and $\rho_m$ and $\dot{\rho}_m$ (determined by Eq.(5) with $n = 3$). The result is differential equation:

$$R^2 d\rho_v + 2\rho_v R dR - a_0 \rho_v R_0 dR = \rho_{m0} R_0 \left(\frac{R_0}{R}\right)^2 dR; \qquad a_0 \equiv \frac{3}{4}\frac{\lambda_\pi^3}{l_{Pl}^2 R_0} \approx 1.68 \qquad (8)$$

having as solution

$$\rho_v = \rho_{v0}\left(\frac{R_0}{R}\right)^2\left\{\left(1+\frac{1}{a_0}\frac{\rho_{m0}}{\rho_{v0}}\right)e^{a_0\left(1-\frac{R_0}{R}\right)} - \frac{1}{a_0}\frac{\rho_{m0}}{\rho_{v0}}\right\} \quad (9)$$

Now, from Eq.(6), the acceleration of expansion of the Universe may be expressed as

$$\ddot{R} = \pi G \frac{\lambda_\pi^3}{l_{Pl}^2}\rho_v = \frac{3H^2}{8}\frac{\lambda_\pi^3}{l_{Pl}^2}\Omega_v \quad (10)$$

or, using Eq.(9):

$$\ddot{R} = \frac{3}{8}\frac{\lambda_\pi^3}{l_{Pl}^2}\Omega_{v0}H_0^2\left(\frac{R_0}{R}\right)^2\left\{\left(1+\frac{1}{a_0}\frac{\Omega_{m0}}{\Omega_{v0}}\right)e^{a_0\left(1-\frac{R_0}{R}\right)} - \frac{1}{a_0}\frac{\Omega_{m0}}{\Omega_{v0}}\right\} \quad (11)$$

Let's note that in the last two equations I have introduced usual dimensionless density parameters $\Omega_m$ and $\Omega_v$, instead of $\rho_m$ and $\rho_v$.

The Eq.(11) shows that there is a critical value of the scale factor $R_{crit}$, determined by

$$\frac{R_0}{R_{crit}} = 1 + \frac{1}{a_0}\ln\frac{a_0\Omega_{v0}+\Omega_{m0}}{\Omega_{m0}} \approx 2.1 \quad (12)$$

so that $\ddot{R} < 0$ when $R < R_{crit}$ and $\ddot{R} > 0$ when $R > R_{crit}$. Hence the accelerated expansion of the Universe has started when Universe was about half of its present size (i.e. $R_{crit} \approx 0.48 R_0$); what is earlier than prediction of standard cosmology: $R_{crit} = (\Omega_{mo}/2\Omega_{v0})^{1/3} R_0 \approx 0.54 R_0$.

However, for the late time Universe ($R \gg R_0$) there is a dramatic difference between acceleration (11) decreasing as $(R_0/R)^2$ and the result of standard cosmology predicting an acceleration $\ddot{R}_{sc}$ increasing linearly with the scale factor i.e. $\ddot{R}_{sc} = \Omega_{v0}H_0^2 R$. Consequently, in the late-time Universe, cosmological scale factor is a linear function of time; while the standard cosmology predicts an exponential function.

## 4. Revision of Dirac's relation for mass of a pion

A simple transformation of Eq.(10) gives

$$m_\pi^3 = \frac{\hbar^2}{cG}H\left\{3\pi^3 c\frac{H\Omega_v}{\ddot{R}}\right\} \quad (13)$$

The incomplete relation (13), without the term in brackets, i.e. proportionality

$$m_\pi^3 \sim \frac{\hbar^2}{cG}H \quad (14)$$

was known to Dirac [3] and Weinberg [4], but there are problems with relation (14). The Hubble parameter (and hence the right-hand side of relation (14)) is a function of the age of the Universe; while the left-hand side of the same relation should be a constant. In fact, in order to get the right mass of a pion, we are forced to choose $H = H_0$ in relation (14), and even so the left-hand side is about one order of magnitude greater than the right one. In order to save relation (14) as a fundamental one, Dirac has suggested [3] that the ratio $H/R$ must stay constant with time; hence introducing a varying gravitational "constant" not supported by observations [4]. The alternative with constant ratio $H/c$, introducing a varying speed of light was considered as well [5]. My position is quite different: relation (14) is considered as an incomplete relation which must be completed in an appropriate way. In fact, without invoking varying "constants" my conjecture (6) leads to the "missing" dimensionless term in brackets, having needed numerical value close to 12, and assuring that the right hand side of Eq.(13) does not change with the expansion of the Universe.

Of course, Eq.(13) corresponds only to the conjectured equation of state (7). The corresponding modification of relation (14) for the standard equation of state, relating dark energy with cosmological constant is given in Ref.[1]

## 5. Comments

If the conjecture (7) is correct, the pressure of the perfect fluid modelling dark energy is function of three variables: dark energy density, matter density and the size of the Universe. It shouldn't be a surprise. The standard equation of state (4) assuming that pressure depends only on dark energy density is presumably an oversimplification neglecting a possible impact of matter on the physical vacuum. It seems plausible to me that matter acts as an external gravitational field "inducing" a certain pressure in the physical vacuum. Hence pressure should have two components: "induced" pressure and "internal" pressure; what is incorporated in conjecture (7).

The fundamental relation (6) allows an amusing interpretation, as energy density of a gas of virtual gravitational dipoles, having the same number density as virtual pions! To see it, we need two things. First, let's remember that in the framework of Quantum Field Theories, physical vacuum is considered as a "kingdom" of virtual particle-antiparticle pairs. Second, let's assume that gravitational mass (gravitational charge) of particle and antiparticle has opposite sign and hence there is gravitational repulsion between particles and antiparticles. This conjecture of gravitational repulsion between particles and antiparticles has two major consequences relevant for relation (6).:

First, it is immediately clear that a virtual particle-antiparticle pair is a system with zero gravitational mass and such a cancelation of gravitational masses might be ground for an eventual solution of the cosmological constant problem. By the way, a similar cancellation of the opposite electric charges of particle and antiparticle in a virtual pair, leads to the zero density of the electric charge of the vacuum.

Second, a virtual pair may be considered as a gravitational dipole with gravitational dipole moment $\vec{p} = m\vec{d}$; $\vec{d}$ may be defined as a vector directed from the antiparticle to the particle, and presenting the distance between them. Distance $d$ should be of the order of a Compton wavelength i.e. $d \sim \lambda_m = \hbar/mc$ and hence $p \sim \hbar/c$. As virtual dipoles are subject of accelerated expansion of the Universe, a virtual gravitational dipole should have energy related to this acceleration: $\varepsilon \sim \hbar \ddot{R}/c$. Now energy density of the gas of virtual dipoles should be $\rho_{de} \sim N_0 \hbar \ddot{R}/c$ where $N_0$ presents number of dipoles per unit volume. It is essentially conjecture (6) if we identify $N_0$ with the number of virtual pions per unit volume (i.e. $N_0 \sim 1/\lambda_\pi^3$).

A final note: relation (6) may be rewritten as

$$\rho_{de} = \frac{A}{\lambda_\pi^3} k_B T_v ; \quad k_B T_v = \frac{1}{2\pi} \frac{\hbar}{c} \ddot{R} \qquad (15)$$

($k_B$ is Boltzmann constant) and $T_v$ may be interpreted as temperature of the vacuum related to accelerated expansion of the Universe. Hence, the second of relations (15) attributes to the physical vacuum a universal temperature having the same value for all observers; let's note that this temperature has the same mathematical form as the famous Unruh [6] and Hawking [7] temperature. It is evident from equations (11) and (15) that vacuum temperature decreases with the expansion of the Universe, approaching (but never reaching) the absolute zero.

In conclusion, it seems that equation of state (7) deserves further more detailed considerations, as a serous alternative, to the standard equation of state (4).